\documentclass[a4paper]{report}
\usepackage[utf8]{inputenc}
\usepackage[T1]{fontenc}
\usepackage{RJournal}
\usepackage{amsmath,amssymb,array}
\usepackage{booktabs}

\usepackage{subfigure}

\newcommand{\rcosmo}{\pkg{rcosmo}\,}

\begin{document}

\sectionhead{Contributed research article}
\volume{XX}
\volnumber{YY}
\year{20ZZ}
\month{AAAA}

\begin{article}
\title{\pkg{rcosmo}: R Package for Analysis of Spherical,
HEALPix and Cosmological Data}
\author{Daniel Fryer, Ming Li, Andriy Olenko}

\maketitle

\abstract{
	The analysis of spatial observations on a sphere is important in areas such as geosciences, physics and
	embryo research,  just to name a few. The purpose of the package \rcosmo is to conduct efficient information processing, visualisation, manipulation and spatial statistical analysis of Cosmic Microwave Background (CMB) radiation and other spherical data. The package was developed for spherical data stored in the Hierarchical Equal Area isoLatitude Pixelation (Healpix) representation. \rcosmo has more than 100 different functions. Most of them initially were developed for CMB, but also can be used for other spherical data as \rcosmo contains tools for transforming spherical data in cartesian and geographic coordinates into the HEALPix representation.  We give a general description of the package and illustrate some important functionalities and benchmarks.}

\section{Introduction}
Directional statistics deals with data observed at a set of spatial directions, which are usually positioned on the surface of the unit sphere or star-shaped random particles.  Spherical methods are important research tools in geospatial, biological, palaeomagnetic and astrostatistical analysis,  just to name a few. The books \citep{fisher, mardia} provide comprehensive overviews of  classical practical spherical statistical methods. Various stochastic and statistical inference modelling issues are covered in \citep{Yadrenko, marinucci_peccati_2011}.

The CRAN Task View \ctv{Spatial} shows several packages for Earth-referenced data mapping and analysis. All currently available R packages for spherical data can be classified in three broad groups. 

The first group provides various functions for working with geographic and spherical coordinate systems and their visualizations. Probably the most commonly used R package to represent spatial maps and data is \CRANpkg{sp} \citep{sp}. It includes tools for  spatial selection, referencing and plotting spatial data as maps. It has comprehensive hierarchical classes and methods for spatial 2d and 3d data. Another example, \CRANpkg{sphereplot} \citep{sphereplot}, uses \CRANpkg{rgl} \citep{rgl} to create 3d visualizations in the spherical coordinate system.  The functions \code{car2sph} and \code{sph2car} implement transformations between Cartesian and spherical coordinates. The package \CRANpkg{geosphere} \citep{geosphere} includes  functions for computing distances, directions and areas for geographic coordinates. 

The second group covers various numerical procedures that can be useful for  spherical approximations and computations. For example, \CRANpkg{SpherWave} \citep{CircNNTSR}  is developed to implement the spherical wavelets and conduct the multiresolution analysis on the sphere.  Functions for numerical integration over high-dimensional spheres and balls are provided in the package \CRANpkg{SphericalCubature} \citep{SphericalCubature}.

The third group provides statistical tools for spherical data analysis. In this group, the most commonly used packages are \CRANpkg{RandomFields} \citep{RandomFields} and \CRANpkg{geoR} \citep{geoR}.  These packages provide a number of tools for model estimation, spatial inference, simulation, kriging, spectral and covariance analyses. Most of their underlying models are for 2d or 3d data, but some additional spherical models are listed for future developments.   The package \CRANpkg{Directional} \citep{Directional}  has functions for von Mises-Fisher kernel density estimation, discriminant and regression analysis on the sphere. The package \CRANpkg{gensphere} \citep{gensphere} implements  multivariate spherical distributions. \CRANpkg{CircNNTSR} \citep{CircNNTSR} provides functions for statistical analysis of spherical data by means of non-negative trigonometric sums. The package \CRANpkg{VecStatGraphs3D} \citep{VecStatGraphs3D} conducts statistical analysis on 3d vectors. It includes estimation of parameters and some spherical test. Another example is the package \CRANpkg{sm} \citep{sm} for spherical regression analysis and non-parametric density estimation.

There are also several R packages developed for spherical data in astronomy\footnote{see the list in  \url{https://asaip.psu.edu/forums/software-forum/459833927}}. For example, \CRANpkg{cosmoFns} \citep{cosmoFns} and \CRANpkg{CRAC} \citep{CRAC} implement functions to compute spherical geometric quantities useful for cosmological research. The package \CRANpkg{FITSio} \citep{FITSio}  reads and writes files in one of standard astronomical formats, FITS (Flexible Image Transport System). \CRANpkg{spider}  \citep{spider}  includes functions for all-sky grid/scatter plots under various astronomical coordinate systems (equatorial, ecliptic, galactic).  The package \CRANpkg{astro} \citep{astro} provides functions for basic cosmological statistics and FITS file manipulations.
 
In recent years the spatial analysis and theory of spherical data have been strongly motivated by the studies on the Cosmic Microwave Background (CMB) radiation data collected by NASA and European Space Agency missions COBE, WMAP and Planck and usually stored in the Hierarchical Equal Area isoLatitude Pixelation (HEALPix) format. 
Cosmologists have developed comprehensive Python and MATLAB software packages\footnote{\url{https://healpix.sourceforge.io/},  \url{https://healpy.readthedocs.io}, \url{http://sufoo.c.ooco.jp/program/healpix.html}}  to work with CMB and HEALPix data.  

Although the mentioned before R packages can be used for geographic or spherical coordinate referenced data,  comprehensive and easy to use tools for CMB and HEALPix data are not available in R, which motivated the authors to design the package \CRANpkg{rcosmo} \citep{rcosmo}.

The aims of the package \rcosmo are 
\begin{itemize}
\item to give convenient access and integrated in one package tools for analysis and visualisation of CMB and HEALPix data to the R statistical community;
	\item to develop R functions for models and methods in spherical statistics that are not available in the existing R packages;
	\item to extend familiar R classes to spherical HEALPix data making them cross-compatible and intuitively interactable  with many existing R statistical packages.
\end{itemize}
The HEALPix format has numerous advantages to classical geographic representations of spherical data, see \citep{originalHealpix}. R implementation of computationally expensive statistical and geometrical methods, such as nearest neighbour searches, empirical covariance function estimation, uniform sampling, spectral density estimation, in a way that takes advantage of the HEALPix data structure, could be useful for geostatistics and other applications. It can reduce algorithmic complexity and computational time. Various data processing, manipulation, visualisation and statistical analysis tasks are achieved efficiently in \rcosmo, using optimised C++ code where necessary.  

\section{Basics of CMB data}
In the Standard Cosmological Model, the Cosmic Microwave Background is redshifted microwave frequency light that is believed to have originated around 14 billion years ago.  CMB is the main source of data about the early universe and seeds of future galaxies. Bell Labs physicists Arno Penzias and Robert Wilson received the Nobel prize in physics in 1978 for their famously happenstance discovery of CMB radiation as an inconvenient background ``noise''  during their experimentation with the Holmdel Horn Antenna radio telescope \citep{[66]}.

Over a decade later, NASA's Cosmic Microwave Background Explorer (COBE) satellite mission produced the first detailed full CMB sky map \citep{[70]}. Referred to as the dawn of precision cosmology COBE results provided fine constraints on many cosmological parameters. Particular attention has been paid to the existence of CMB anisotropies and associated non-Gaussianity, usually investigated through the CMB angular power spectrum \citep{[66]}.
The Wilkinson Microwave Anisotropy Probe, was launched in 2001 by NASA and returned more precise measurements of CMB \citep{[69]}. Then, the third and most detailed space mission to date was conducted by the European Space Agency, via the Planck Surveyor satellite \citep{[68]}. The radiation that astronomers detect today forms an expanding spherical surface of radius approximately 46.5 billion light years. The next generation of CMB experiments, CMB-S4, LiteBIRD, and CORE, will consist of highly sensitive telescopes. It is expected that these experiments will provide enormous amount of CMB measurements and maps to nearly the cosmic variance limit.

The term ``CMB data'' refers to a broad range of location tagged quantities describing properties of the CMB. For example, the Infrared Science Archive  by Caltech's Infrared Processing and Analysis Center (IPAC) hosts curated CMB products from the North American Space Agency (NASA)\footnote{hosted at the link \url{http://irsa.ipac.caltech.edu}}. 

To produce CMB maps (see Figure~\ref{window-plot}), the products of the Planck mission data (in the range of frequencies from $30$ to $857$ GHz) are separated from foreground noise using one of the four detailed methods named COMMANDER, NILC, SEVEM and SMICA\footnote{\url{https://wiki.cosmos.esa.int/planckpla2015/index.php/Astrophysical_component_separation}}. These CMB maps are provided at either low resolution ($N_{\text{side}} = 1024$, i.e., $10$ arcmin resolution), or high resolution ($N_{\text{side}} = 2048$, i.e., $5$ arcmin resolution). The maps include temperature intensity and polarisation data, as well as common masks for identifying regions where the reconstructed CMB is untrusted.

Our focus will mostly be on the CMB temperature intensity data. In Planck CMB products, these data are stored as 4-byte floating point binary numbers in  $K_{\text{cmb}}$ defined as the unit in which a black body spectrum at $2.725$ Kelvin (K) is flat with respect to the frequency \citep{[65]}.

\section{Statistical model}
The map of the CMB temperature is usually modelled as a realization of an isotropic Gaussian random field on the unit sphere. This section introduces a statistical model and basic notations of spherical random fields. More details can be found, for example, in the monographs \citep{marinucci_peccati_2011}, \citep{Yadrenko} and the paper \citep{leonenko2018}.

Consider a sphere in the three-dimensional Euclidean space
$
\mathbb{S}^2=\left\{ \mathbf{x}\in \mathbb{R}^{3}:\Vert \mathbf{x}\Vert =1\right\} \subset \mathbb{R}^{3} .$

A spherical random field on a probability space $(\Omega ,\mathcal{F%
},\mathbf{P})$, denoted by%
\begin{equation*}
T=\left\{ T(\theta ,\varphi )=T_{\omega }(\theta ,\varphi ):0\leq \theta
\leq \pi ,\text{ }0\leq \varphi \leq 2\pi ,\ \omega \in \Omega \right\} ,
\end{equation*}%
or $\widetilde{T}=\{\widetilde{T}(\mathbf{x})$ $,$ $\mathbf{x}\in \mathbb{S}^2\},$ is a stochastic function
defined on the sphere $\mathbb{S}^2.$

The field $\widetilde{T}(\mathbf{x})$ is called isotropic  on the sphere $\mathbb{S}^2$ if its mean $ \mathbf{E}T(\theta ,\varphi )=c=constant$
and the covariance function $ \mathbf{E}T(\theta
,\varphi )T(\theta ^{\prime },\varphi ^{\prime })$ depends only on the
angular distance $\Theta =\Theta _{PQ}$ between the points $P=(\theta
,\varphi )$ and $Q=(\theta ^{\prime },\varphi ^{\prime })$ on $\mathbb{S}^2.$

A real-valued second-order mean-square continuous spherical random
field $T$ can be expanded in the series
\begin{equation}
T(\theta ,\varphi )=\sum_{l=0}^{\infty }\sum_{m=-l}^{l}a_{lm}Y_{lm}(\theta
,\varphi ),
\label{field}
\end{equation}
where $\{Y_{lm}(\theta ,\varphi )\}$ represents the complex spherical harmonics and $a_{lm}$ are random variables.

If a random field is isotropic then
\[
\mathbf{E}a_{lm}a_{l^{\prime }m^{\prime }}^{\ast }=\delta _{l}^{l^{\prime
}}\delta _{m}^{m^{\prime }}C_{l},\quad -l\leq m\leq l,\quad -l^{\prime }\leq
m^{\prime }\leq l^{\prime },\quad l\ge 0.
\]%
Thus, $ \mathbf{E}|a_{lm}|^{2}=C_{l},$ $m=0,\pm 1,...,\pm l.$ The series $\left\{ C_{1},C_{2},...,C_{l},...\right\}$ is called the
angular power spectrum of the isotropic random field $T(\theta ,\varphi ).$

The covariance function of the isotropic random fields $T$  has the  following representation
\begin{equation}
\Gamma(\cos\Theta )= \mathbf{E}T(\theta ,\varphi )T(\theta ^{\prime },\varphi
^{\prime})=\frac{1}{4\pi }\sum_{l=0}^{\infty}(2l+1)C_{l}P_{l}(\cos\Theta),
\label{covar}
\end{equation}%
where $P_l(\cdot)$ is the $l$-th Legendre polynomial and $\sum_{l=0}^{\infty }(2l+1)C_{l}<\infty.$

The variogram (semivariogram) function is defined by 
\begin{equation}\label{vario}
\gamma(h) =\Gamma(0) - \Gamma(h).
\end{equation}

In the following we use observations of the field $T$ at HEALPix points. There are two main approaches in the statistical analysis of $T$ that are realised in \rcosmo. First approach directly uses the observations for parameter estimation and hypothesis tests. For example, the classical estimator of the isotropic variogram $\gamma(h)$ takes the form of
\[\hat{\gamma}(h)=\frac{1}{2N_h}\sum_{(\mathbf{x}_1,\mathbf{x}_2)\in N_h} \left(T(\mathbf{x}_1) - T(\mathbf{x}_2)\right)^2,\] 
where $N_h$ is the set of the spherical location pairs at the geodesic distance $h.$ 
The second approach is spectrum-based. Initially, the  estimates $\hat{a}_{lm}$ are computed by inverting (\ref{field}). Then, empirical covariance functions and variograms can be obtained by substituting the estimated values of the angular power spectrum \[\hat{C}_{l}=\frac{1}{2l+1}\sum_{m=-l}^l|\hat{a}_{lm}|^2\] in equations \ref{covar} and \ref{vario}.

\section{\pkg{rcosmo} package}
The current version of the \rcosmo package can be installed from CRAN. A development release is available from GitHub (\url{https://github.com/frycast/rcosmo}).  

The package offers various tools for

\begin{itemize}
\item Handling and manipulating of Cosmic Microwave Background radiation and other spherical data,
\item  Working with Hierarchical Equal Area isoLatitude Pixelation of a sphere (Healpix),
\item Spherical geometry,
\item Various statistical analysis of CMB and spherical data,
\item Visualisation of HEALPix data.
\end{itemize}

Most of \rcosmo features were developed for CMB, but it can also be used for other spherical data.

\begin{figure}[ht]
	\centering
	\includegraphics[trim={0cm 0cm 0cm 0cm},clip,,width=1\textwidth]{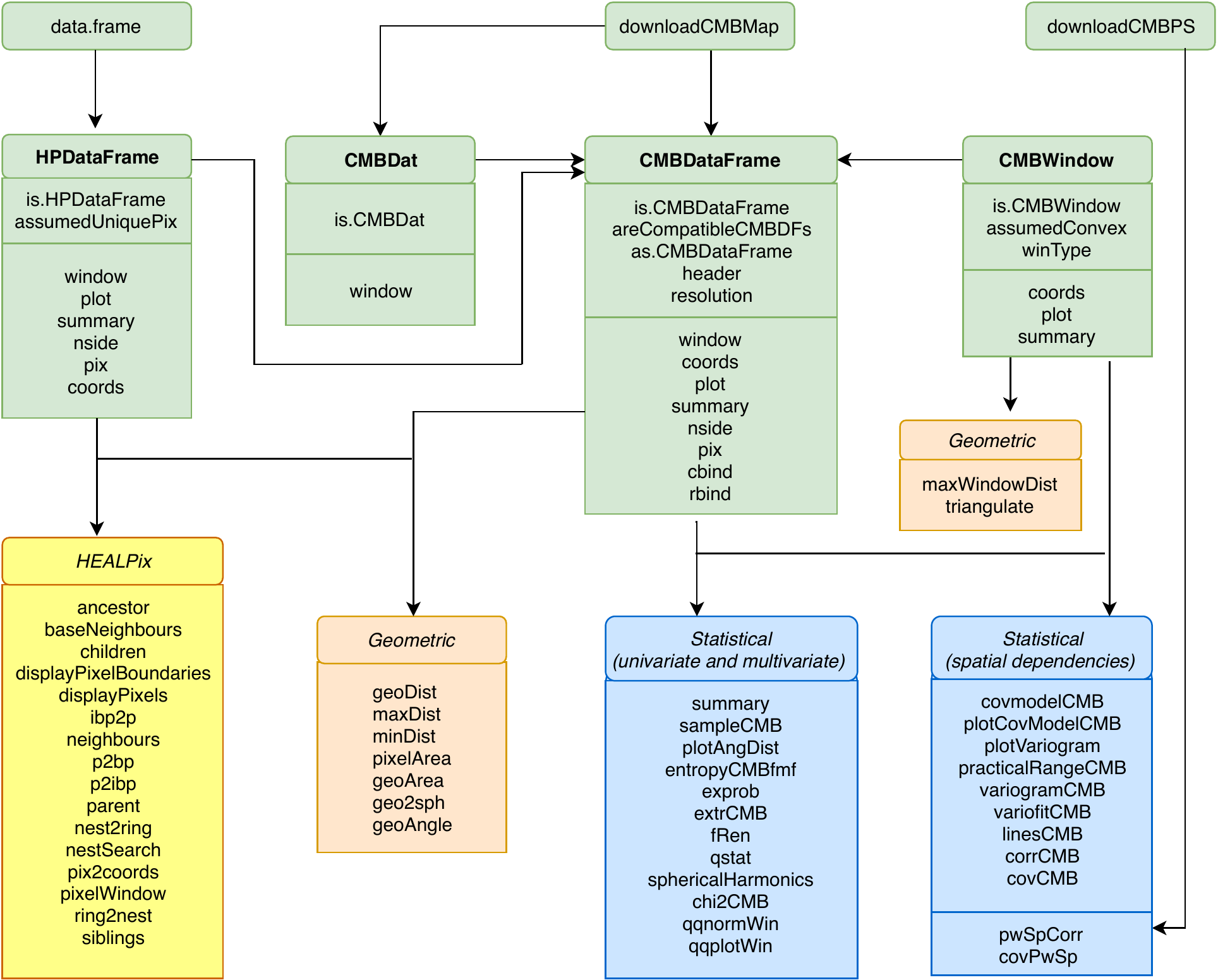}
	\caption{\rcosmo main structure and core functions.}
	\label{map}
\end{figure}

The package has more than 100 different functions. Figure~\ref{map} shows the core functions available in \rcosmo and some typical data analysis flow sequences.

Rather than attempting a systematic description of each functions, the remainder of this paper shows broad classes of methods implemented in \rcosmo with particular examples of core functions. A reproducible version of the code in this paper is available in the folder "Research materials" from the  website~\url{https://sites.google.com/site/olenkoandriy/}.

\section{Visualisation tools}
Standard Python and MATLAB tools for CMB and HEALPix visualization use the Mollweide projection of the unit sphere to the 2d plane.  This is an equal-area projection, but it distorts spherical angles and distances. In contrast \rcosmo employs the \samp{OpenGL} powered 3d visualization device system \pkg{rgl} for R to allow 3d interactive plots of data on the unit sphere. The generic \code{plot} function produces interactive 3d vector graphics that may be easily exported to a HTML document.  

Some examples of using \code{plot.CMBDataFrame} are given in Figures~\ref{window-plot} and \ref{annulus-plot}. For an example of using \code{plot.HPDataFrame} see Figure~\ref{world-plot} and  examples of using \code{plot.CMBWindow} are shown in Figures~\ref{window-plot} and \ref{triang}.  For better visualization some figures produced by the R code in this paper were rotated and zoomed in before including in the article.

By default, the Planck colour scale is applied to CMB intensity data for plotting\footnote{Colour scale is available here: https://github.com/zonca/paperplots/tree/master/data}. Interactive visualisations of spherical data are a focus point of \rcosmo, and additional features such as automatic plot legends, alternative colour scales, and greater configurability are planned for future releases. In addition, \rcosmo provides a variety of 2d plot functionality to support visualisation of statistical analysis results and some additional 3d plot functionality for demonstrating HEALPix pixel properties.

\section{\pkg{rcosmo} classes}

Four R classes have been developed to support HEALPix data representation and analysis in the package \rcosmo:  \code{"CMBDat"}, \code{"CMBDataFrame"}, \code{"HPDataFrame"} and \code{"CMBWindow"}. First three are main parent classes of objects to store spherical data. The class \code{"CMBWindow"} is used to choose observation windows.

\begin{itemize}
	\item The function \code{CMBDat} creates objects of class \code{"CMBDat"}. \code{CMBDat} objects are lists containing header information, other metadata and a data slot. Data slots may include standard information about CMB intensity (I), polarisation (Q, U), PMASK and TMASK. It also may contain a \code{mmap} object that points to the CMB map data table in a FITS file. As for standard data frames new data slots can be created to store other types of spherical data.
	\item   The function \code{CMBDataFrame} creates objects of class \code{"CMBDataFrame"}. These class is a special modification of 	\code{"data.frame"} that also carries metadata about, e.g., the HEALPix ordering scheme, coordinate system,   and nside parameter (i.e., the resolution of the  HEALPix grid that is being used). Each  row of a \code{CMBDataFrame} is associated with a unique HEALPix pixel index.
	\item  The class \code{"HPDataFrames"} is a type of \code{"data.frame"} designed to carry data located on the unit sphere.  Unlike \code{"CMBDataFrame"}, \code{"HPDataFrames"} may have repeated pixel indices. It allows to store multiple data points falling within a given pixel in different rows of \code{HPDataFrame} objects.
	\item The function \code{CMBWindow} creates objects of class \code{"CMBWindow"}. These objects are polygons, spherical discs, or their compliments, unions and intersections.
\end{itemize}

As the main \rcosmo data classes are special modifications of \code{"data.frame"} it means that spatial objects produced by \rcosmo can be subsequently processed by other R packages/functions that work with standard data frames. The \code{rbind} and \code{cbind} generics that work with the \code{"data.frame"} class have been customised in \rcosmo to preserve the validity of \code{CMBDataFrame} objects.

\section{Getting data into rcosmo}

In this section, we demonstrate how to import CMB data in the typical case of a full sky map stored as a FITS file. Such maps can be sourced from the NASA/IPAC Infrared Science Archive\footnote{hosted by Caltech at the link \url{http://irsa.ipac.caltech.edu/data/Planck/release_2/}}.

The function \code{downloadCMBMap} can be used to download Planck CMB maps. One can specify the type of map (\samp{COMMANDER}, \samp{NILC}, \samp{SEVEM} or \samp{SMICA}), the resolution ($N_{\text{side}} = 1024$ or $2048$), and save it in a working directory with a specified file name.

The map \file{COM\_CMB\_IQU-smica\_1024\_R2.02\_full.fits} used in most of the examples in this paper is a FITS file of approximate size 200 megabytes. This map has the resolution $N_{\text{side}} = 1024$, so it contains $N_{\text{pix}} = 12\times1024^2 = 12582912$ pixels, each having its own intensity $I$, polarisation $Q,U,$ temperature mask value $T_{\text{mask}} \in \{0,1\}$ and polarisation mask value $P_{\text{mask}} \in \{0,1\}$. 

After downloading the map with \code{downloadCMBMap} and applying the function \code{CMBDataFrame}, we obtain an object of class \code{"CMBDataFrame"}. 
\begin{example}
> filename <- "CMB_map_smica1024.fits"
> downloadCMBMap(foreground = "smica", nside = 1024, filename)
> sky<-CMBDataFrame(filename) 
> str(sky)
Classes ‘CMBDataFrame’ and 'data.frame':	12582912 obs. of  5 variables:
$ I    : num  -9.20e-05 -8.04e-05 -8.99e-05 -7.71e-05 -7.01e-05 ...
$ Q    : num  6.47e-08 -9.19e-09 7.36e-08 5.45e-09 -7.10e-08 ...
$ U    : num  -6.57e-07 -6.94e-07 -6.85e-07 -7.18e-07 -7.30e-07 ...
$ TMASK: int  0 0 0 0 0 0 0 0 0 0 ...
$ PMASK: int  0 0 0 0 0 0 0 0 0 0 ...
- attr(*, "nside")= int 1024
- attr(*, "ordering")= chr "nested"
- attr(*, "resolution")= num 10
...
\end{example}

An alternative to the above act of reading the entire map into memory is to take a random sample of points on the sphere. This is achieved without reading the entire map into R memory. Simple random sampling in \rcosmo will be discussed further under the section on statistical functions.

\begin{example}
> set.seed(0); s <- 2e6; 
> cmb_sample <- CMBDataFrame(filename, sample.size = s, include.m = T, include.p = T)
> cmb_sample
A CMBDataFrame
# A tibble: 200,000 x 5
             I             Q            U TMASK PMASK
         <dbl>         <dbl>        <dbl> <int> <int>
 1 -0.0000517   0.0000000328 -0.000000613     0     0
 2 -0.0000206  -0.000000182  -0.000000714     0     0
...
10 -0.00000943 -0.00000169   -0.00000127      0     0
# ... with 199,990 more rows
\end{example}

%
%
%
%

\section{Use of memory mapping}

The standard library for reading data from FITS files is a collection of C and FORTRAN subroutines called \samp{CFITSIO}. In R, the package \pkg{FITSio} provides an interface to \samp{CFITSIO} \citep{FITSio}. However, importing a full sky CMB map with approximately 12 million intensity samples from a FITS file using \pkg{FITSio} took too long when development of \rcosmo began. We were able to reduce the necessary run time to under 4 seconds with the \rcosmo function \code{CMBDat}. \rcosmo still substantially outperforms the last version of \pkg{FITSio}.  In the following example we used the CMB map with \samp{SMICA} foreground separation and $N_{side} = 2048$ (having approximately 50 million intensity samples)  to test it on a modern laptop\footnote{Laptop specifications: Microsoft Surface Laptop with 7th Gen Intel Core m3 (i5) processor; 8GB LPDDR3 SDRAM (1866MHz)}.

\begin{example}
> filename1 <- "CMB_map_smica2048.fits"
> downloadCMBMap(foreground = "smica", nside = 2048, filename1)
> system.time(sky <- CMBDataFrame(filename1))
user  system elapsed 
0.85    0.28    1.28

> system.time(fits <- FITSio::readFITS(filename1))
user   system   elapsed
454.64   3.33    460.13
\end{example}

The approach used in \rcosmo is based on a novel application of the \pkg{mmap} package by Jeffrey Ryan \citep{mmap}. The package \pkg{mmap} is a highly optimised interface to \samp{POSIX mmap} and Windows \samp{MapViewOfFile}.  Using \pkg{mmap} in \rcosmo required an update of \pkg{mmap} to support big-Endian byte order. The current version of \pkg{mmap} allows us to import data from a FITS binary table very efficiently, one row at a time, using a C struct data type. Ideally, for a typical \rcosmo user, the details of using \pkg{mmap} are abstracted away while the user constructs and interacts with \code{"CMBDataFrame"} objects.

Another use of \pkg{mmap} in \rcosmo concerns the elimination of the need to read a large full sky CMB map into R memory. Often it is unmanageable to read the entire contents of a FITS file. For example, it may not be possible to obtain sufficiently large blocks of continguous memory from the operating system when importing more than a few hundred megabytes of data as a numeric vector. In \rcosmo, integration of \pkg{mmap} allows one to maintain a C-style pointer at a particular byte-offset to the target binary file (e.g., the FITS file), so that data can be read into R memory on command from this offset. In the following example, only rows $1,2,4,7$ and $11$ are read into memory from the file.

\begin{example}
> v <- c(1,2,4,7,11)
> sky <- CMBDataFrame(filename, spix = v, include.p = T, include.m = T,
 coords = "spherical")
> sky
A CMBDataFrame
# A tibble: 5 x 7
  theta   phi          I        Q            U TMASK PMASK
  <dbl> <dbl>      <dbl>    <dbl>        <dbl> <int> <int>
1  1.57 0.785 -0.0000920  6.47e-8 -0.000000657     0     0
2  1.57 0.786 -0.0000804 -9.19e-9 -0.000000694     0     0
3  1.57 0.785 -0.0000771  5.45e-9 -0.000000718     0     0
4  1.57 0.786 -0.0000663 -5.81e-8 -0.000000751     0     0
5  1.57 0.783 -0.0000836  7.60e-8 -0.000000697     0     0

> pix(sky)
[1]  1  2  4  7 11
\end{example}

The next section discusses the HEALPix data structure. HEALPix ordering schemes can be used to map coordinates on the sphere to row indices in a FITS binary table. Combining this feature with the technique in the above example allows \rcosmo to efficiently import random samples of data from a variety of geometric regions of the sphere without ever having to import the entire CMB map. This is particularly useful on larger maps and will become increasingly important in future as advances in cosmology allow for higher resolution CMB maps to be produced. 

\section{Introduction to HEALPix} \label{healpix-intro}

Present generation Cosmic Microwave Background experiments produce data with up to 5 arcminutes resolution on the sphere. For a full-sky map, this amounts to approximately 50 million pixels, each describing distinct location, intensity, polarisation and other attributes. The statistical analysis of such massive datasets, and associated discretisation of functions on the sphere, can involve prohibitive computational complexity  and non-adequate sampling in the absence of an appropriate data structure. The Hierarchical Equal Area isoLatitude Pixelation is a geometric structure designed to meet this demand using a self-similar refinable mesh. It is currently the most widely used pixelation for storing and analysing CMB data \citep{originalHealpix}.

The package \rcosmo provides various tools to visualize and work with the HEALPix structure.

\begin{figure}[htbp]
	\centering
	\subfigure[HEALPix base pixel boundaries]{\includegraphics[scale = 0.45]{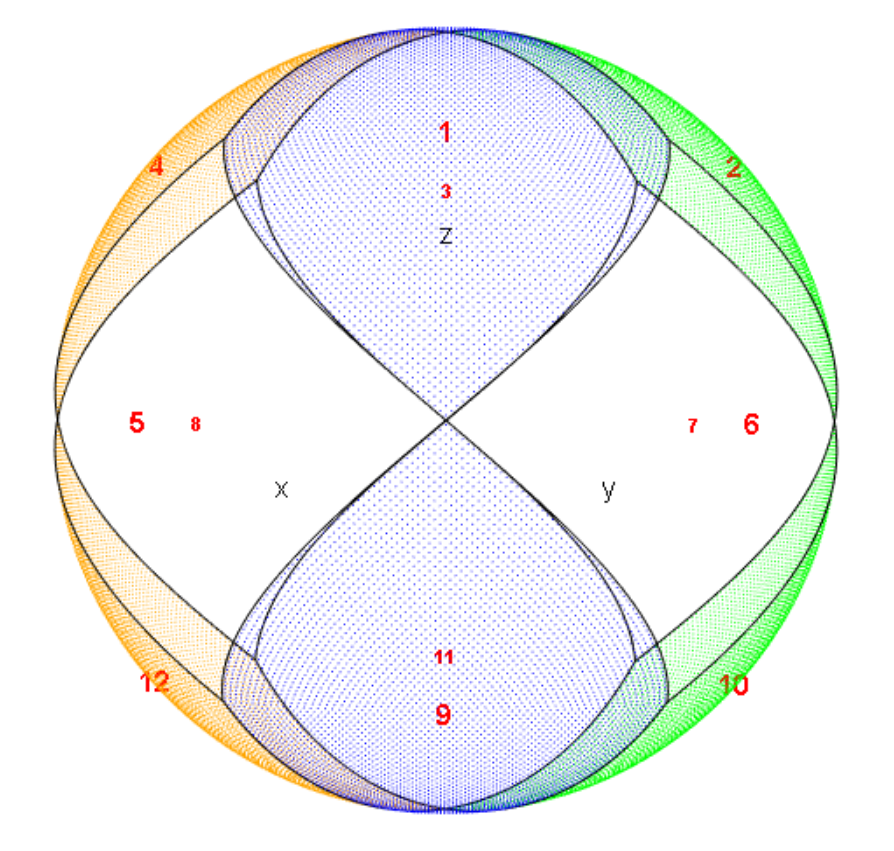}\label{fig-base-pixels}}\hspace{8mm}
	\subfigure[HEALPix nested ordering]{\includegraphics[scale = 0.45]{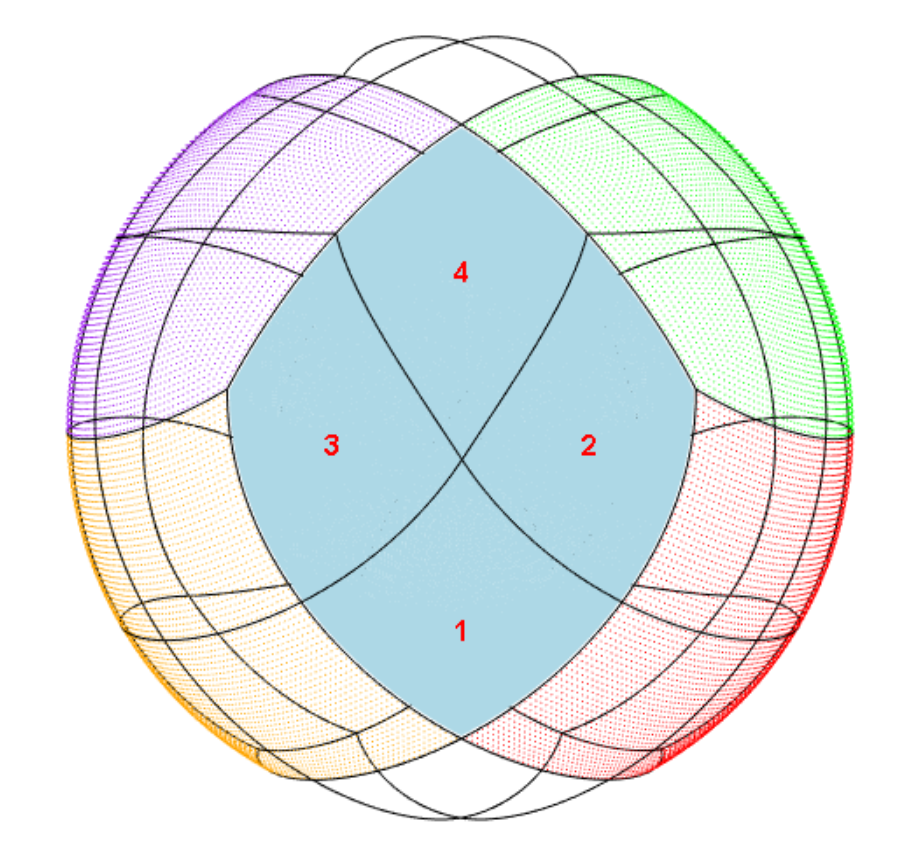}\label{fig-nside2-pixels}}
	\caption{(a): HEALPix base pixel boundary visualisation. The 12 base pixels are labelled 1 to 12; (b):~HEALPix nested ordering visualisation at $N_{\text{side}} = 2.$ Pixels 1, 2, 3 and 4 (labelled) all fall within base pixel 1 (coloured solid).}
\end{figure}

HEALPix initially divides the sphere into 12 equiareal \dfn{base pixels}. To visualise these with \rcosmo, we can first generate a \code{CMBDataFrame} at some low resolution (e.g, $N_{\text{side}} = 64$) and then take three separate window subsets in the pixels that we intend to colour, as shown in Figure \ref{fig-base-pixels}). Note that, while all HEALPix pixels are 4-sided, their edges are not geodesics, i.e., they are not spherical quadrillaterals \citep{[39]}. 
\begin{example}
> ns <- 64; rand <- rnorm(12 * ns ^ 2)
> cmbdf <- CMBDataFrame(nside = ns, I = rand, ordering = "nested")
> w1 <- window(cmbdf, in.pixels = c(1,9))
> w2 <- window(cmbdf, in.pixels = c(2,10))
> w3 <- window(cmbdf, in.pixels = c(4,12))
> plot(w1, col = "blue", back.col = "white")
> plot(w2, col = "green", add = TRUE)
> plot(w3, col = "orange", add = TRUE)
> displayPixelBoundaries(nside = 1, ord = "nested", incl.lab = 1:12, col = "red")
\end{example}
Each of the 12 base pixels can be further subdivided into 4 equiareal 4-sided pixels. For a demonstration, we can create another window subset based on a higher resolution \code{CMBDataFrame} and display the outputs in Figure \ref{fig-nside2-pixels}.
\begin{example}
> ns <- 256; rand <- rnorm(12 * ns ^ 2)
> cmbdf <- CMBDataFrame(ns = ns, I = rand, ordering = "nested")
> w <- window(cmbdf, in.pixels = 1)
> plot(w, col = "light blue", back.col = "white" , add = TRUE, size = 1.2)
> plot(window(cmbdf, in.pixels = 2), col = "green" , add = TRUE)
> plot(window(cmbdf, in.pixels = 4), col = "purple", add = TRUE)
> plot(window(cmbdf, in.pixels = 5), col = "orange", add = TRUE)
> plot(window(cmbdf, in.pixels = 6), col = "red"   , add = TRUE)
> displayPixelBoundaries(nside = 2, ord = "nested", incl.lab = 1:4, col = "black")
\end{example}

This process of subdivision can be repeated until a desired resolution is achieved. At the required resolution, the number of edge segments per base pixel edge is referred to as the $N_{\text{side}}$ parameter, and satisfies $N_{\text{pix}} = 12N_{\text{side}}^2$, where we use  $N_{\text{pix}}$ to denote the total number of pixels.

\begin{figure}[htbp]
\begin{center}
\includegraphics[scale = 0.4]{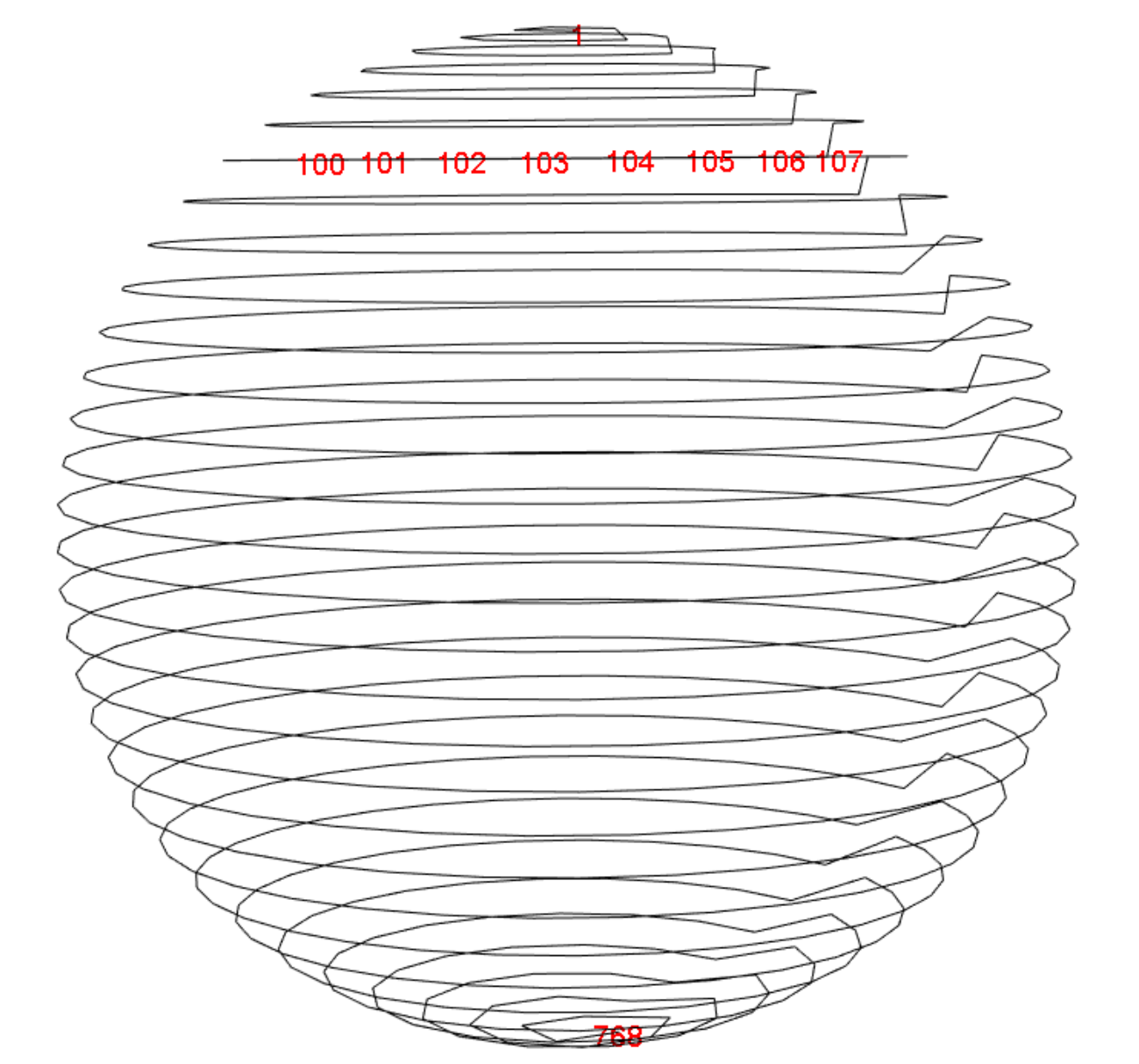}
\end{center}\vspace{-5mm}
\caption[HEALPix ring ordering.]{HEALPix ring ordering scheme visualisation. The black line traces in order through pixel centers from 1 to $N_{\text{pix}} = 768$. The locations of pixels 100 to 107 are labelled.}
\label{fig-ring-pixels}
\end{figure}

At a given $N_{\text{side}}$, the HEALPix representation provides a bijection from the first $12N_{\text{side}}^2$ natural numbers $P$ to a set of locations $L$ on the unit sphere. We refer to $P$ as the set of \dfn{pixel indices} and $L$ as the set of \dfn{pixel centers}. For assigning the pixel indices to the pixel centers there are two approaches, known respectively as the "ring" and "nested" \dfn{ordering schemes}. The nested scheme is demonstrated with the numbering in Figure \ref{fig-nside2-pixels}. The ring scheme is demonstrated in Figure~\ref{fig-ring-pixels}. 

\begin{example}
> cmbdf <- CMBDataFrame(nside = 8, ordering = "ring")
> plot(cmbdf, type = 'l', col = "black", back.col = "white")
> tolabel <- c(1,100:107,768)
> plot(cmbdf[tolabel,], labels = tolabel, col = "red", add = TRUE)
\end{example}

Regardless of the choice of ordering scheme, all HEALPix pixel centers lie on $4N_{\text{side}} - 1$ rings of constant latitude. This feature facilitates fast discrete spherical harmonic transforms, since the associated Legendre functions need only be evaluated once per isolatitude ring of pixel centers \citep{originalHealpix}.

Many tasks gain or lose efficiency with the choice of ordering scheme. For example, nearest neighbour searches are best conducted in the nested scheme \citep{originalHealpix}. As such, every object of class \code{"CMBDataFrame"} or \code{"HPDataFrame"} has an attribute named \code{ordering} to indicate which of the two schemes is being used. This allows \rcosmo functions to choose the most efficient scheme for each task, performing any necessary conversions with the internal functions \code{nest2ring} and \code{ring2nest}. 

\section{HEALPix functions}

For working directly with HEALPix properties there are a number of \rcosmo functions. Some core functions are shown in Figure~\ref{map} in the yellow colour, other are internal and some were exported. Broadly, there are the following categories:

\begin{itemize}
\item Working with ordering schemes, 
\item Navigating the nested ordering hierarchy,
\item Geometric functions involving pixel indices, 
\item Visualising the HEALPix structure.
\end{itemize}

The main ordering functions include converting between two ordering schemes and getting information about a type of ordering (\code{ordering} generic and internals \code{nest2ring} and \code{ring2nest}). For example, the \code{ordering} generic function is useful for getting and setting the ordering attribute of a \code{CMBDataFrame} or \code{HPDataFrame}.  
\begin{example*}
> sky <- CMBDataFrame(nside = 2, ordering = "ring"); ordering(sky)
[1] "ring"
> ordering(sky) <- "nested"; ordering(sky)
[1] "nested"
\end{example*}

\rcosmo functions for navigating the HEALPix structure provide various tools to investigate local neighbourhoods of specific pixels and relative positions of pixels at different levels of the nested ordering hierarchy, see, for example, \code{ancestor}, \code{pixelWindow}, \code{neighbours}, etc. Since the nested ordering is self-similar, many of these functions  are resolution independent. 

For example, the $k^{\text{th}}$ \dfn{ancestor} of a pixel index $p$ at resolution $j := \log_2(N_{\text{side}})$ is the pixel index $p_a$ to which $p$ belongs, $k$ steps up the hierarchy (i.e., at resolution $j - k$). It turns out that $p_a$ is a function $p_a = f(p,k)$ that is independent of $j$.  The following example produces the ancestors of pixel $p = 10^3$, for $k = 1,2,\ldots, 5,$ using the internal function \code{ancestor}.

\begin{example*}
> ancestor(1e3, 1:5)
[1] 250  63  16   4   1
\end{example*}

\begin{figure}[htbp]
\begin{center}
\includegraphics[scale = 0.3]{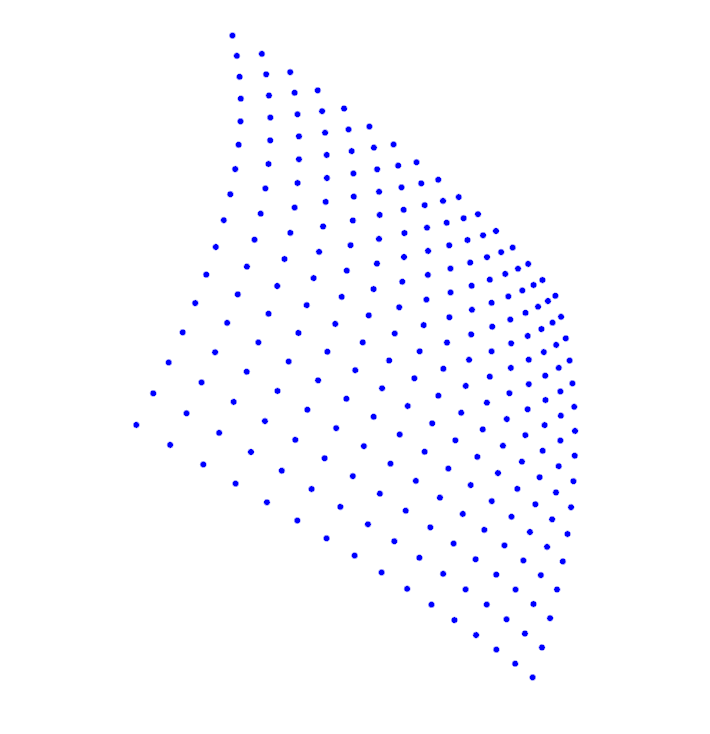}
\end{center}\vspace{-5mm}
\caption[Pixel center plot]{All pixel centers (at resolution 5), within pixel 1 (at resolution 1)}
\label{pixel-plot}
\end{figure}

A function that is not resolution independent is \code{pixelWindow}. In the following code, \code{pixelWindow} retrieves all pixels at resolution $j_2 = 5$ that lie within pixel $p = 1$, specified at resolution $j_1 = 1.$ The result is shown in Figure~\ref{pixel-plot}.

\begin{example*}
p <- 1; j1 <- 1; j2 <- 5
P <- pixelWindow(j1 = j1, j2 = j2, pix.j1 =  p)
displayPixels(spix = P, j = j2, plot.j = j2)
\end{example*}

The group of \rcosmo functions that includes \code{pix2coords}, \code{pixelArea}, \code{nestSearch}, etc.,  computes spherical geometric properties in relation to by pixel indices. For example, the \code{nestSearch} function searches a pixel closest to a point in 3d space. It uses an algorithm that achieves a high level of efficiency using the nested hierarchy. A comparison, via \CRANpkg{microbenchmark} \citep{microbenchmark}, with a basic linear search algorithm which we call \code{geoDistSearch} reveals the following.

\begin{example}
> library(microbenchmark)
> geoDistSearch <- function(target, nside, spix) {
    xyz <- pix2coords(nside = nside, spix = spix)
    dists <- geoDist(xyz, target)
    return(xyz[which.min(dists),])}
 
> t <- data.frame(x = 0.6, y = 0.8, z = 0)
> nside <- 16; p <- 1:(12*(nside)^2)
> mb <- microbenchmark::microbenchmark(
    geoDistSearch(target = t, nside = nside, spix = p),
    nestSearch(target = t, nside = nside))
> summary(mb)$median[1]/summary(mb)$median[2]
[1] 219.8153
\end{example}

From the above, it was observed that \code{nestSearch} was over $200$ times faster than \code{geoDistSearch} at finding the closest pixel center at $N_{\text{side}}  = 16$ to the point $(x,y,z) = (0.6, 0.8, 0).$ With \code{nestSearch}, the closest pixel to a target point is found by checking only $12+4\log_2(N_{\text{side}})$ pixels rather than $12N_{\text{side}}^2.$   When $N_{\text{side}} = 2048$, only $12+4\log_2(2048) = 56$ pixels must be checked from over 50 million. 

%


\section{Subsetting and combining spherical regions}
\label{CMB:subsetting}
\rcosmo functions for selecting and visualizing spherical regions can be broadly divided in the following groups:
\begin{itemize}
	\item Creating basic \code{CMBWindow} objects (polygons or spherical discs),
	\item Combining  different sub-regions by using compliments, unions and intersections to create a new \code{CMBWindow} object,
  \item Plotting a region with boundary and inside points,
    \item Extracting data from a given \code{CMBDataFrame} restricted to a \code{CMBWindow} region. 
\end{itemize}

Class \code{CMBWindow} is designed to carry geometrical information describing the interior or exterior of spherical figures (polygons, spherical discs (caps), and their complements, unions and intersections). The polygons can be non-convex, though \code{CMBWindow} carries a boolean attribute \code{assumedConvex} that should be set to \code{TRUE}, if the polygon is known in advance to be convex. In this case special methods that decrease computation times are applied. 

\begin{figure}[htbp]
\centering
\subfigure[The Dragon]{\includegraphics[scale = 0.3]{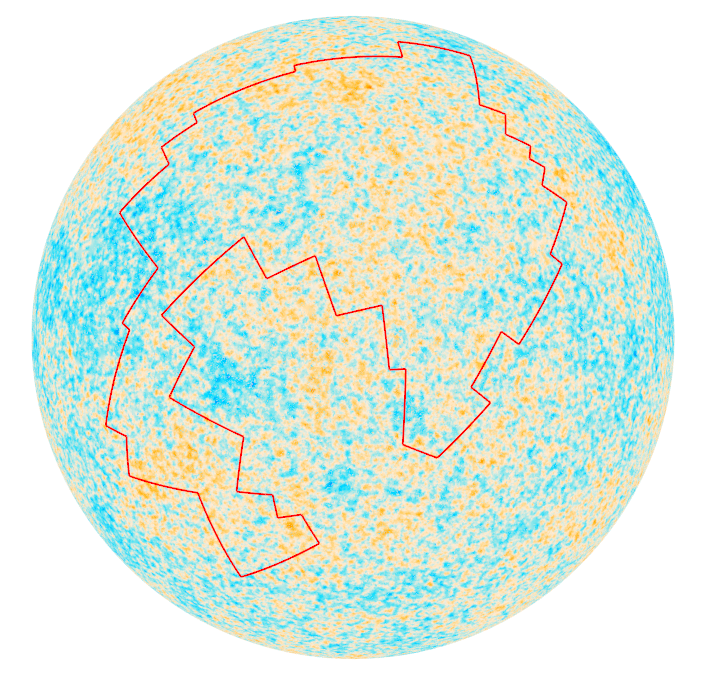}\label{fig-ploygon1}}\hspace{10mm}
\subfigure[The Scorpion]{\includegraphics[scale = 0.29]{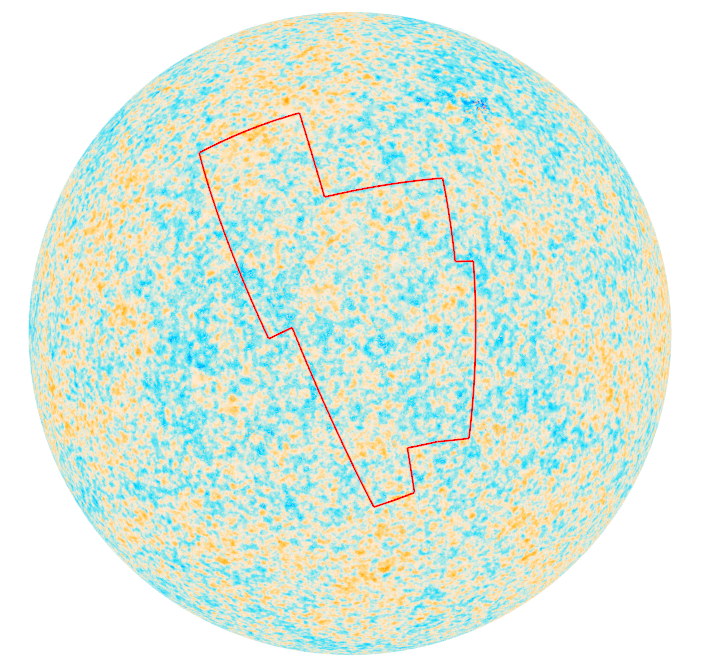}\label{fig-ploygon2}}
\caption[CMBWindow boundaries for polygon and disc.]{Boundary visualisation of polygon CMBWindow objects, plotted against $10^5$ CMB intensities.}
\label{window-plot}
\end{figure}

A \code{CMBWindow} object can be created using the \code{CMBWindow} function. The code below illustrates the creation of two \code{CMBWindow}s that correspond to the Dragon and Scorpion constellations. Files of constellation boundaries\footnote{available at \url{https://www.iau.org/public/themes/constellations/}} include coordinates of spherical polygons vertices that correspond to each constellation. The function \code{hms2deg} converts celestial coordinates (hours, minutes, seconds) to the degrees format. Then "phi" and "theta" columns of the \code{data.frame} \code{CB} are used to create CMBWindow objects. 
To inspect these \code{CMBWindow} objects using interactive 3D graphics, we can pass them to the generic plot function. Below, we also plot the CMB map as a background. The resulting plot is displayed in Figure \ref{window-plot}.

\begin{example}
> download.file("https://www.iau.org/static/public/constellations/txt/dra.txt",
  "bound1.txt")
> x1 <- readLines("bound1.txt")
> x1 <- gsub("\\|", " ", x1)
> Constellation_Boundary1 <- read.table(text = x1,col.names=c("H","M","S","D","Con_N"))
> download.file("https://www.iau.org/static/public/constellations/txt/sgr.txt",
  "bound2.txt")
> x2 <- readLines("bound2.txt")
> x2 <- gsub("\\|", " ", x2)
> Constellation_Boundary2 <- read.table(text = x2,col.names=c("H","M","S","D","Con_N"))
 
> CB0 <- Constellation_Boundary1
> deg <- celestial::hms2deg(CB0[,1],CB0[,2],CB0[,3])
> CB1 <- data.frame(pi*deg/180, pi*CB0[,4]/90)
> colnames(CB1) <- c("phi","theta")
> polygon1 <- CMBWindow(phi = CB1$phi, theta = CB1$theta)
> plot(cmb_sample, back.col = "white")
> plot(polygon1, lwd=2)
\end{example}

After repeating the steps above for  the Scorpion constellation we obtain and the second \code{CMBWindow} object \code{polygon2} in Figure \ref{window-plot}.


Note that for the \code{CMBWindow} polygons defined above, entire polygons lie within any one hemisphere of $\mathbb{S}_2.$ To obtain \code{CMBWindow} objects that occupy more than one hemisphere, we can specify a polygon or disc exterior (complement in $\mathbb{S}_2$) using the \code{set.minus = TRUE} parameter. For example, the following command gives the exterior of a spherical cap with a base radius 0.5.

\begin{example}
> d.exterior <- CMBWindow(theta = pi/2, phi = 0, r = 0.5, set.minus = TRUE)
\end{example}

\begin{figure}[htbp]
\begin{center}
\includegraphics[scale = 0.8]{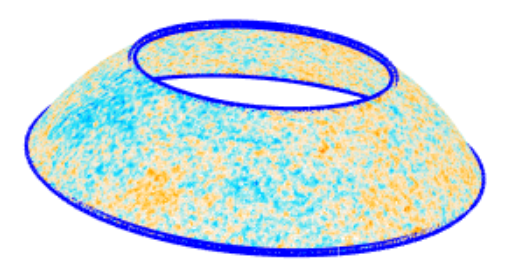}
\end{center}\vspace{-5mm}
\caption[Annular subregion of a CMBDataFrame.]{CMB intensity data extracted from an \code{CMBDataFrame} object by the \code{window} function. }
\label{annulus-plot}
\end{figure}

To specify more complicated regions, we can combine multiple \code{CMBWindow} objects into a \code{list}. For example, the following command results in the list containing d.exterior and the interior of a spherical disc (cap) of base radius 1 (disc's radius is computed on the sphere surface), which is a spherical segment shown in Figure~\ref{annulus-plot}.

\begin{example}
> wins <- list(d.exterior, CMBWindow(theta = pi/2, phi = 0, r = 1))
\end{example}

By passing \code{CMBWindow} objects to the \code{window} function, one can extract data from a \code{CMBDataFrame} or \code{mmap} object. Below, using the above spherical window \code{wins} the \code{CMBDataFrame} named \code{sky.annulus} is created and plotted in Figure~\ref{annulus-plot}.

\begin{example}
> df <- CMBDataFrame(filename)
> sky.annulus <- window(df, new.window = wins)
> plot(sky.annulus, back.col = "white")
> plot(wins[[1]], lwd = 5, col="blue"); plot(wins[[2]], lwd = 5, col="blue")
\end{example}

\section{Spherical geometry functions}

Several basic tools for spherical geometry are implemented in \rcosmo:

\begin{itemize}
	\item Converting between different coordinate systems on the sphere,
	\item Computing geodesic distances between points and windows,
	\item Calculating  spherical angles,
	\item Computing areas of spherical figures,
	\item Triangulating spherical polygons.
\end{itemize}

The  currently implemented core geometric functions are shown in Figure~\ref{map} in the orange colour. Some other spherical geometric tools are specified for the HEALPix representation or \code{CMBWindows} and are shown in the green colour.

For example, the functions \code{geoArea} computes the area of a figure on the unit sphere that is encompassed by its pixels.

\begin{example}
> geoArea(sky.annulus)
[1] 2.11917
\end{example}

Another example is the function \code{maxDist} that computes the maximum geodesic distance either between all points in a \code{data.frame} pairwise, or between all points in a \code{data.frame} and a target point.
\begin{example}
> p <- c(0,0,1)
> maxDist(sky.annulus, p)
[1] 2.570185
\end{example}

\begin{figure}[htbp]
 \begin{center}
 \includegraphics[scale = 0.5,trim={0 0 0 0.5cm},clip]{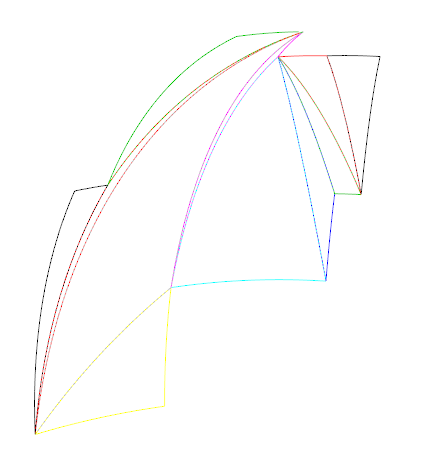}
 \end{center}\vspace{-5mm}
 \caption[Triangulation]{Triangulation of a spherical polygon.}
 \label{triang}
 \end{figure}

Various geometric problems require triangulations of spherical polygons. For a polygonal \code{CMBWindow} the function \code{triangulate} produces a set of spherical triangles with pairwise disjoint interiors and the union equals to the original polygon. For example, Figure~\ref{triang} shows a triangulation of the Dragon constellation spherical polygon.

\begin{example}
> win1 <- triangulate(polygon2)
> for (i in 1:11) {plot(win1[[i]],  col=i)}
\end{example}

\section{Statistical functions} \label{statistical-functions}

In this section we overview core statistical functions implemented in \rcosmo. The package provides various tools for statistical analysis of spherical data that can be broadly divided in the following types:
\begin{itemize}
	\item Spherical random sampling,
	\item Univariate spherical statistics and plots,
	\item Multivariate statistics for data from different \code{CMBWindows},
	\item Measures of spatial dependencies.
\end{itemize}

The main currently implemented statistical functions are shown in Figure~\ref{map} in the blue colour. Below we provide few examples of functions for each type.

\subsection{Random sampling}

An immediate advantage of equiareal HEALPix pixel sizes is that simple random sampling is not regionally dependent \citep{originalHealpix}. That is, a simple random sample of pixel indices produces an approximately uniform sample of locations on the sphere.

To get a simple random sample from a \code{CMBDataFrame} one can use the function \code{sampleCMB}. This function returns a \code{CMBDataFrame} which size equals to the function's parameter \code{sample.size}. This new object has rows that comprise a simple random sample of the rows from the input CMBDataFrame.

\begin{example}
> set.seed(0)
> sampleCMB(df, sample.size = 3)
A CMBDataFrame
# A tibble: 3 x 5
           I           Q            U TMASK PMASK
       <dbl>       <dbl>        <dbl> <int> <int>
1 -0.0000868  0.00000314  0.000000471     1     1
2  0.0000341  0.00000201 -0.00000308      1     1
3 -0.000104  -0.00000161 -0.00000550      1     1
\end{example}

\subsection{Univariate  spherical statistics and plots}
There are several methods in \rcosmo for statistical analysis and visualisation of CMB temperature intensity data. For example, function \code{summary} produces a CMBDataFrame summary that includes information about window's type and area, total area covered by observations, and the main statistics for the intensity data in the spherical window.

\begin{example}
> summary(sky.annulus)
=============================  CMBDataFrame Object  ============================
Number of CMBWindows:  2 
+-----------------------------+
|                             |
|   Window type: minus.disc   |
|   Window area: 11.7972      |
|                             |
+-----------------------------+

+-------------------------+
|                         |
|   Window type: disc     |
|   Window area: 2.8884   |
|                         |
+-------------------------+
METHOD  = 'smica   '           / Separation method                               
Total area covered by all pixels:  2.11917 
~~~~~~~~~~~~~~~~~~~~~~~~~~~~~~~~~~~~~~~~~~~~~~~~~~~~~~~~~~~~~~~~~~~~~~~~~~~~~~~~
Intensity quartiles
Min.    1st Qu.     Median       Mean    3rd Qu.       Max. 
-7.609e-04 -6.908e-05 -6.208e-07 -1.487e-06  6.737e-05  7.697e-04 
================================================================================
\end{example}

 The function \code{entropyCMB} returns an estimated entropy for specified  column intensities and \code{CMBWindow}.
\begin{example}
> entropyCMB(cmbdf = df, win = d.exterior, intensities = "I")
[1] 2.13523
\end{example}

 The first Minkowski functional \code{fmf} returns an area of the spherical region, where the intensities are above of the specified threshold level $\alpha.$
\begin{example}
> fmf(cmbdf = sky.annulus, alpha = 0, intensities = "I")
[1] 1.054269
 \end{example}

\begin{figure}[htbp]
 \begin{center}
 \includegraphics[scale = 0.5, trim={0cm 0cm 0 1.5cm},clip]{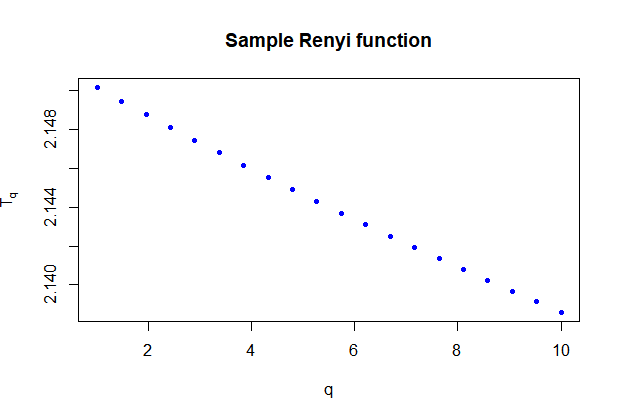}
 \end{center}\vspace{-5mm}
 \caption[Sample Renyi function]{Sample Renyi function $\hat{T}(q)$ of \code{sky.annulus} on [1.01,10].}
 \label{Renyi}
 \end{figure}

\code{fRen} computes values of the sample R\'{e}nyi function $\hat{T}(q)$ on the grid of $N$ uniformly spaced points in the interval [q.min, q.max]. The R\'{e}nyi function describes fractal properties of random fields and can be used for testing departures from Gaussianity, see \citep{leonenko2013}. For example,  Figure~\ref{Renyi}  shows that $\hat{T}(q)$ is an approximate strait line for for the data in \code{CMBWindow} \code{sky.annulus}. Thus,  there is not enough evidence to reject Gaussianity of the random field based on the data and their sample R\'{e}nyi function in this \code{CMBWindow}.
\begin{example}
> Tq <- fRen(cmbdf = sky.annulus, q.min = 1.01, q.max = 10, N = 20, intensities = "I")
> plot(Tq[,1], Tq[,2], ylab =expression(T[q]), xlab = "q", main = "Sample Renyi 
  function", pch = 20, col = "blue")
 \end{example}

\begin{figure}[htbp]
 \begin{center}
 \includegraphics[scale = 0.7]{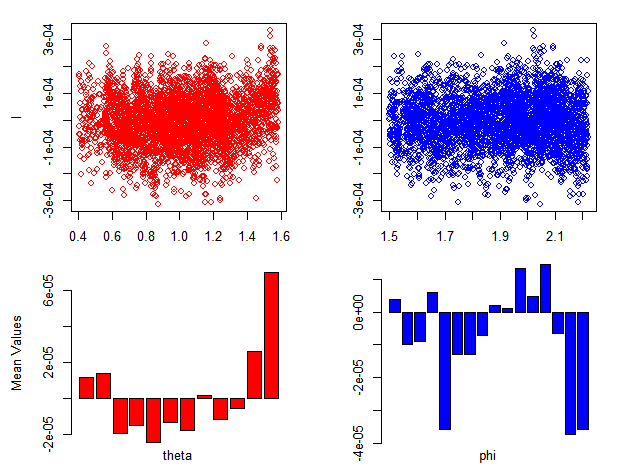}
 \end{center}\vspace{-5mm}
 \caption[Marginal distributions]{Distributions of temperature versus $\theta$ and $\phi$ angles for the Scorpion constellation region.}
 \label{AngDis}
 \end{figure}

The function \code{plotAngDis} helps to visualise the marginal distributions of temperature intensities versus $\theta$ and $\phi$ angles.  It produces scatterplots and  barplots of the corresponding means computed over bins, see Figure~\ref{AngDis}.  

\begin{example}
> df1 <- sampleCMB(df, sample.size = 100000)
> cmbdf.win <- window(df1, new.window = polygon2)
> plotAngDis(cmbdf.win, intensities = "I")
 \end{example}

\subsection{Multivariate statistics for data from different CMBWindows.}

There are several \rcosmo functions for comparison of data from two or more \code{CMBWindows}. For example, the function \code{qqplotWin} is a modification of the standard \code{qqplot} to  produces a QQ plot of quantiles of observations in two \code{CMBWindows} against each other for a specified CMBDataFrame column.

The example below shows that the distributions of temperatures in the Dragon and  Scorpion constellations are similar.
\begin{example}
> qqplotWin(df1, polygon1, polygon2)
 \end{example}
\begin{figure}[ht]
 \begin{center}
 	\includegraphics[scale = 0.5,trim={1cm 1cm 0 1cm},clip]{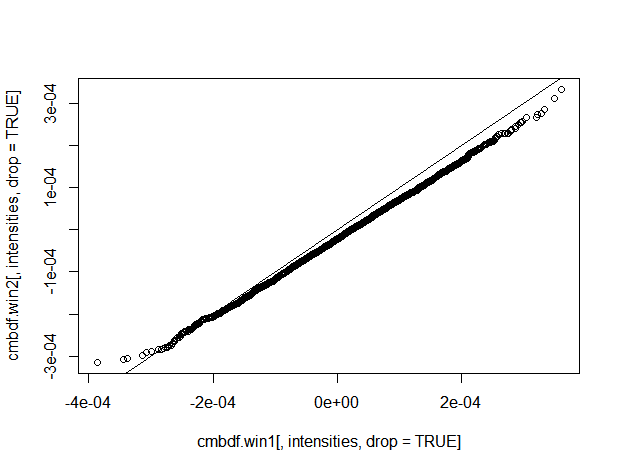}
 \end{center}\vspace{-5mm}
 \caption[QQ plot]{QQ plot of observations in the Dragon vs  Scorpion constellation regions.}
 \label{qqplot}
\end{figure}

The function \code{qstatq} can be used to measure spatial stratified heterogeneity in a list of \code{CMBWindows}. It takes values in $[0, 1],$ where  0 corresponds to no spatial stratified heterogeneity, 1 means a perfect heterogeneity case. For example, the results below shows that there is not enough evidence for spatial stratified heterogeneity, i.e. the value of the temperature intensities are not different in these two \code{CMBWindows}.
\begin{example}
> lw <- list(polygon1, polygon2)
> qstat(df1, lw)
[1] 0.01136971
\end{example}

\subsection{Investigating spatial dependencies}

This section presents some of \rcosmo tools for the analysis of spatial dependencies in spherical data.

 As the geodesic and  Euclidean distances are different,  covariance functions on $\mathbb{R}^{3}$ can not be used directly for $\mathbb{S}^2.$ The package implemented several parametric models of covariance functions (\ref{covar}) on the sphere, see theoretical foundations in \citep{gneiting2013}. \rcosmo uses the package \CRANpkg{geoR} and extends its list of general spatial models and some functions to the spherical case. Currently available choices of covariance models are  \code{matern,  exponential, spherical,  powered.exponential,  cauchy,}  \code{gencauchy,}  \code{pure.nugget,  askey,} \code{c2wendland,  c4wendland,  sinepower,} and \code{multiquadric.} The default option is \code{matern}.

The function \code{covmodelCMB} computes values of theoretical covariance functions given the separation distance of locations.
The function returns the value of the covariance $\Gamma(h)$ at the geodesic distance $h.$ The covariance model uses the general form $\Gamma(h) =\sigma^2\cdot \rho(h/\psi),$ where the variance $\sigma^2$ and the range $\psi$ are vertical and horizontal scaling parameters respectively.

For example, the following command computes the value of the Askey covariance function with the parameters $\sigma^2=1,$ $\psi = \pi,$ and $\kappa = 4$ at the geodesic distance $h = \pi/4.$ 

\begin{example}
> covmodelCMB(pi/4, cov.pars = c(1, pi), kappa = 4, cov.model = "askey" )
[1] 0.3164062
\end{example}

The command \code{plotcovmodelCMBPlot} is designed to produce quick plots of theoretical covariance functions.
\begin{example}
> plotcovmodelCMB("askey", phi = pi/4, to  = pi/2, kappa = 4)
\end{example}
\begin{figure}[h]
	\begin{center}
		\includegraphics[scale = 0.5,trim={0 0cm 0 1.5cm},clip]{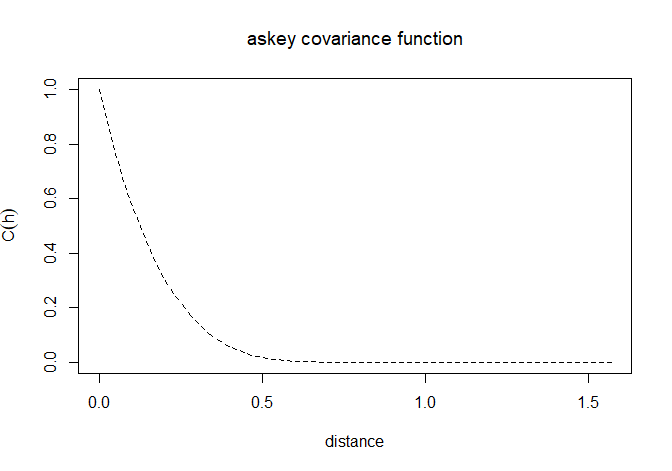}
	\end{center}\vspace{-5mm}
\caption{Plot of the Askey covariance function for the parameters $\sigma^2=1,$ $\psi = \pi/4,$ and $\kappa = 4.$}
\end{figure}

If a random fields is isotropic its covariance function depends only on a geodesic distance between locations. In this case the function \code{covCMB} can be used to compute an empirical covariance function for intensity data in a \code{CMBDataFrame} or \code{data.frame}.  Output is given up to a maximal geodesic distance \code{max.dist}, which can be chosen between 0 and $\pi$ (by default equals $\pi$). 

\begin{example}
> df1 <- sampleCMB(df, sample.size = 100000)
> Cov <- covCMB(df1, max.dist = 0.03, num.bins = 10)
> Cov$v
[1] 1.038704e-08 6.958892e-09 4.237914e-09 3.116515e-09
[5] 2.674736e-09 2.357278e-09 2.375547e-09 2.373873e-09
[9] 2.323980e-09 2.248074e-09 2.206426e-09	
\end{example}

Obtained estimated covariance values can be visualised using the command 
\begin{example}
> plot(Cov)
\end{example}
\begin{figure}[htbp]
	\begin{center}
		\includegraphics[scale = 0.5,trim={0 0.5cm 0 2cm},clip]{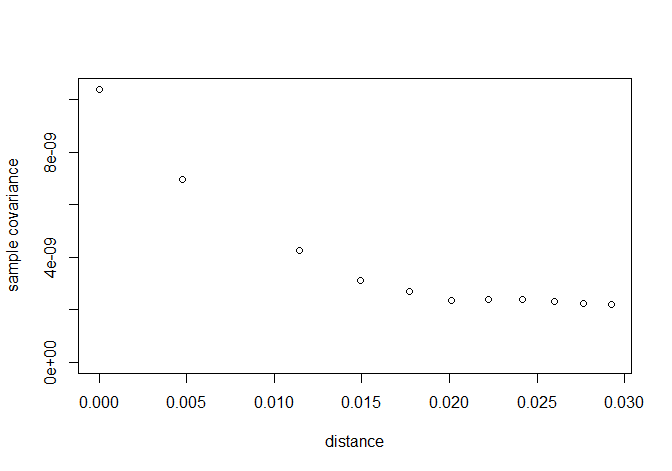}
	\end{center}\vspace{-5mm}
\caption{Plot of the empirical covariance function for \code{max.dist} = 0.03.}
\end{figure}

The function \code{variofitCMB} estimates parameters of variogram models (see equation (\ref{vario}) for the link between covariance and variogram functions) by fitting a parametric model from the list \code{covmodelCMB} to a sample variogram estimated by the function \code{variogramCMB}. This function is built on and extends \code{variofit} from the package \CRANpkg{geoR} to specific \rcosmo covariance models on the sphere.

In the example below the Matern variogram is fitted to the empirical variogram on the interval $[0,0.1]$ using the ordinary least squares method.
\begin{example}
> varcmb <- variogramCMB(df1, max.dist = 0.1, num.bins = 30)
> ols <- variofitCMB(varcmb,  fix.nug=FALSE, wei="equal", cov.model= "matern")
> plot(varcmb)
> lines(ols, lty=2)
\end{example}
\begin{figure}[htbp]
	\begin{center}
		\includegraphics[scale = 0.5,trim={0 0.5cm 0 2cm},clip]{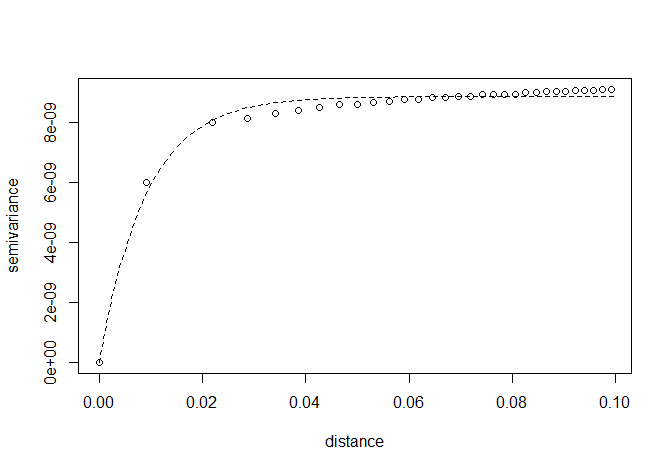}
	\end{center}\vspace{-5mm}
	\caption{Plots of the empirical and fitted variograms.}
\end{figure}

The package also has  tools to work with angular power spectra of spherical random fields.
The CMB power spectrum data are freely available from the section "Cosmology products" of the Planck Legacy Archive\footnote{hosted at the link \url{https://pla.esac.esa.int/pla/}}. They can be easily downloaded in a ready-to-use \rcosmo format by the function \code{downloadCMBPS}. 

\begin{figure}[htbp]
	\begin{center}
		\includegraphics[scale = 0.5,trim={0 0.5cm 0 2cm},clip]{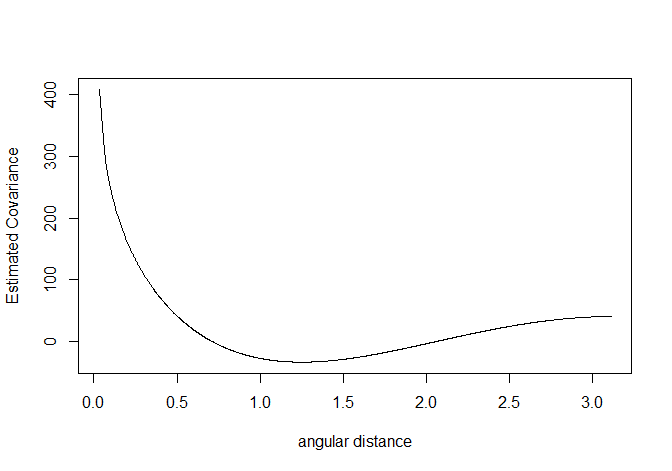}
	\end{center}
\caption{Plot the covariance estimate using CMB power spectrum.}
\label{spec}
\end{figure}

The function \code{covPwSp} uses values of an angular power spectra to provide a covariance estimate by equation (\ref{covar}). As the argument of the covariance function in equation (\ref{covar}) is $\cos\Theta$ the following code uses the inverse transformation \code{acos} to plot the covariance estimate as a function of angular distances in Figure~\ref{spec}.

\begin{example}
> COM_PowerSpectra <- downloadCMBPS(link=1)
> Cov_est <- covPwSp(COM_PowerSpectra[,1:2], 2000)
> plot(acos(Cov_est[,1]), Cov_est[,2], type ="l", xlab ="angular distance", 
  ylab ="Estimated Covariance")
\end{example}

\section{Converting other spherical data to HEALPix format}
While the HEALPix  is the main representation in cosmological applications there are numerous spherical data, for example, in geosciences, that use different coordinate systems and spherical formats. This example shows how non-HEALPix spherical data can be converted to the HEALPix format for \rcosmo analysis.

A \code{HPDataFrame} is suitable for storing data that is located on a sphere, but has not been preprocessed to suit HEALPix structured storage. There are various ways to assign HEALPix pixel indices to the rows of a \code{HPDataFrame}. This example presents the way when a desired resolution is specified in advance. Then the \code{HPDataFrame} constructor automatically assigns pixel indices based on coordinates provided. A row is assigned the pixel index of its closest pixel center. \rcosmo also has the option of automatic computing of a required resolution for  given data to assign data locations to unique pixels.

\begin{figure}[htbp]
\begin{center}
\includegraphics[scale = 0.5]{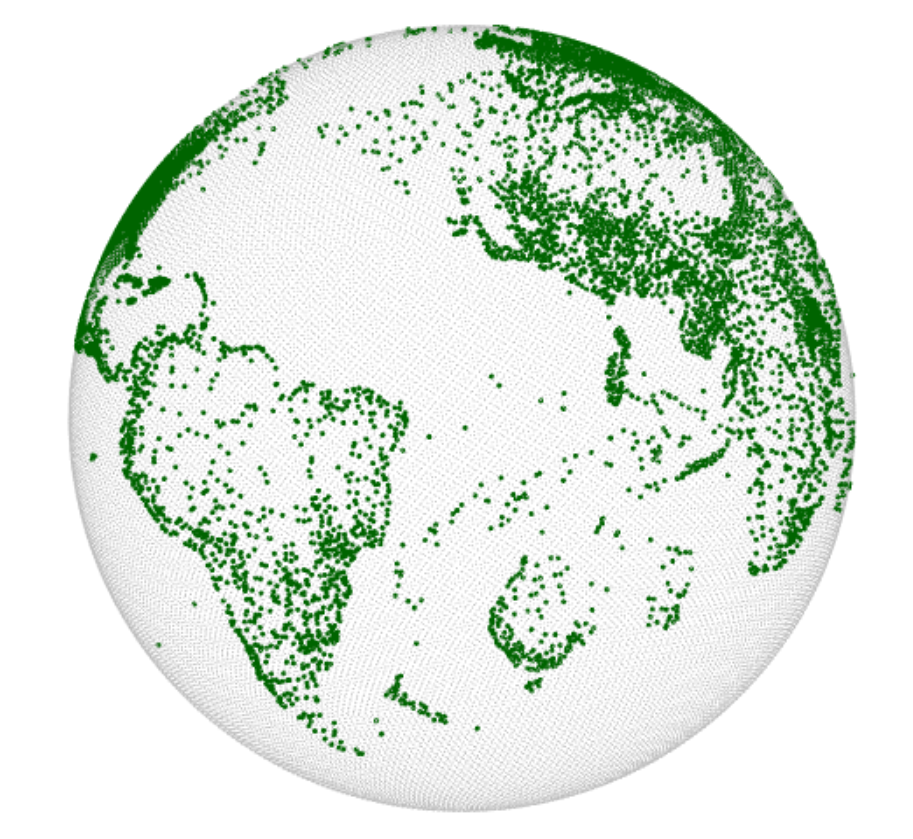}
\end{center}\vspace{-5mm}
\caption{Plot of world's large cities and towns at the resolution \code{nside} = 1024.}
\label{world-plot}
\end{figure}

As an example we use the database with the latitude and longitude of over 13 thousand world's large cities and towns available from the World Cities Database\footnote{hosted at the link \url{https://simplemaps.com/data/world-cities}}. First, cities latitudes and longitudes in degrees are converted to spherical coordinates $(\theta,\phi)$  in radians. Then , we create and visualize \code{HPDataFrame} at the resolution \code{nside} = 1024, see Figure~\ref{world-plot}.
\begin{example}
> worldcities <- read.csv("worldcities.csv")
> sph <- geo2sph(data.frame(lon = pi/180*worldcities$lng, lat = pi/180*worldcities$lat))
> df1 <- data.frame(phi = sph$phi, theta = sph$theta, I = rep(1,nrow(sph)))
> hp <- HPDataFrame(df1, auto.spix = TRUE, nside = 1024)
> plot(hp, size = 3, col = "darkgreen", back.col = "white")
\end{example}
Some other examples of converting directional and star-shaped data into \rcosmo formats will be given in the manuscript that is in preparation.

\section{Summary and future directions}

This article introduces the package \rcosmo for analysis of CMB, HEALPix and other spherical data. The package integrates the HEALPix representation and various spherical geometric and statistical methods in a convenient unified framework. It opens   efficient handling and analysis of HEALPix and CMB data to the R statistical community. \rcosmo also introduces several new spherical statistical models and methods that were not available in  R before. The package can also be very useful for researchers working in geosciences. 

There are several possible extensions that would be useful to the package. Some of them include:
\begin{itemize}
	\item integrating with available Python and C++ HEALPix software,
	\item including new spherical statistical models and methods,
	\item further development of spherical spectral and multifractal methods,
	\item adding new visualisation tools,
	\item improved use of memory mapping for use on extremely high resolution images,
	\item improvements to the user interface based on user feedback.
\end{itemize}

\section{Acknowledgements}
This research was partially supported under the Australian Research Council's Discovery Projects funding scheme (project number  DP160101366). We also would like to thank V.V. Anh, P.Broadbridge, N.Leonenko, I.Sloan, and Y.Wang for their comments on early drafts of \rcosmo and discussions of CMB and spherical statistical methods, and J.Ryan for developing and extending the \pkg{mmap} package.

\bibliography{olenko}

%

\address{Daniel Fryer\\
  Department of Mathematics and Statistics\\
  La Trobe University, VIC, 3086\\
  Australia\\
  ORCiD: 0000-0001-6032-0522\\
  \email{d.fryer@latrobe.edu.au}}

\address{Ming Li\\
  Department of Mathematics and Statistics\\
  La Trobe University, VIC, 3086\\
  Australia\\
  ORCiD: 0000-0002-1218-2804\\
    \email{ming.li@latrobe.edu.au}}

    \address{Andriy Olenko\\
  Department of Mathematics and Statistics\\
  La Trobe University, VIC, 3086\\
  Australia\\
  ORCiD: 0000-0002-0917-7000\\
  \email{a.olenko@latrobe.edu.au}}

\end{article}

\end{document}